%% file: main.tex
\documentclass[12pt]{article}

\usepackage[utf8]{inputenc}
\usepackage{charter}
\usepackage{fullpage}
\usepackage{hyperref}
\usepackage{graphicx}
\usepackage{lipsum}
\usepackage[sort&compress,numbers]{natbib}
\usepackage{hyperref}
\hypersetup{hidelinks}
\usepackage[total={6.5in,9in},top=1in,headsep=0.1in,headheight=1in]{geometry}
\usepackage[utf8]{inputenc}
\usepackage{amssymb}
\usepackage{aas_macros}
\usepackage{ulem}

\newcommand\snowmass{
\begin{center}
  \rule[-0.2in]{\hsize}{0.01in}\\
  \rule{\hsize}{0.01in}\\
  \vskip 0.1in
  Submitted to the Proceedings of the US Community Study\\ 
  on the Future of Particle Physics (Snowmass 2021)\\
  \rule{\hsize}{0.01in}\\
  \rule[+0.2in]{\hsize}{0.01in}\\[-2em]
\end{center}
}

\usepackage[firstpage=true]{background}
\backgroundsetup{contents={\parbox{6.5in}{\snowmass}}, scale=1,placement=top,opacity=1,color=black,position={3.25in,1.2in}}

\usepackage{fancyhdr}
\fancypagestyle{plain}{%
  \fancyhf{}%
  \fancyhead[C]{}
  \fancyfoot[C]{\thepage}
}

\fancypagestyle{empty}{%
  \fancyhf{}%
  \fancyhead[C]{{\it Snowmass2021 Cosmic Frontier: Synergies between DM searches and multiwavelength/multimessenger astrophysics}}
  \fancyfoot[C]{\thepage}
}
\title{Snowmass2021 Cosmic Frontier: Synergies between dark matter searches and multiwavelength/multimessenger astrophysics}
\date{}

\usepackage{authblk}

\author[1,2]{Shin'ichiro Ando}
\author[3]{Sebastian Baum}
\author[4]{Michael Boylan-Kolchin}
\author[5]{Esra Bulbul}
\author[5]{Michael Burgess}
\author[6]{Ilias Cholis}
\author[7]{Philip von Doetinchem}
\author[8]{JiJi Fan}
\author[9]{J. Patrick Harding\thanks{Co-ordinator, jpharding@lanl.gov}}
\author[10,2]{Shunsaku Horiuchi\thanks{Co-ordinator, horiuchi@vt.edu}}
\author[11,12]{Rebecca K. Leane}
\author[1]{Oscar Macias}
\author[13]{Katie Mack}
\author[14,15]{Kohta Murase}
\author[16]{Lina Necib}
\author[17,18]{Ibles Olcina}
\author[19]{Laura Olivera-Nieto}
\author[20]{Jong-Chul Park}
\author[21]{Kerstin Perez}
\author[22,23]{Marco Regis}
\author[24]{Nicholas L. Rodd}
\author[25,26]{Carsten Rott}
\author[27]{Kuver Sinha}
\author[2]{Volodymyr Takhistov}
\author[11]{Yun-Tse Tsai}
\author[28]{Devin Walker\thanks{Co-ordinator, devin.g.walker@dartmouth.edu}}

\affil[1]{GRAPPA Institute, University of Amsterdam, 1098 XH Amsterdam, The Netherlands}
\affil[2]{Kavli IPMU (WPI), UTIAS, The University of Tokyo, Kashiwa, Chiba 277-8583, Japan}
\affil[3]{Stanford Institute for Theoretical Physics, Department of Physics, Stanford University, Stanford, CA 94305, USA}
\affil[4]{Department of Astronomy, The University of Texas at Austin, 2515 Speedway, Stop C1400, Austin, TX, USA}
\affil[5]{Max Planck Institute for Extraterrestrial Physics, 85748 Garching, Germany}
\affil[6]{Department of Physics, Oakland University, Rochester, MI 48309, USA}
\affil[7]{Department of Physics and Astronomy, University of Hawaii at Manoa, Honolulu, HI 96822, USA}
\affil[8]{Department of Physics, Brown University, Providence, RI, 02912, USA}
\affil[9]{Physics Division, Los Alamos National Laboratory, Los Alamos, NM, USA}
\affil[10]{Center for Neutrino Physics, Department of Physics, Virginia Tech, Blacksburg, VA 24061, USA}
\affil[11]{SLAC National Accelerator Laboratory, Stanford University, Stanford, CA 94039, USA}
\affil[12]{Kavli Institute for Particle Astrophysics and Cosmology, Stanford University, Stanford, CA 94039, USA}
\affil[13]{Physics Department, North Carolina State University, Raleigh, NC 27695, USA}
\affil[14]{The Pennsylvania State University, University Park, PA 16802, USA}
\affil[15]{Center for Gravitational Physics, Yukawa Institute for Theoretical Physics, Kyoto University, Sakyo, Kyoto 606-8502, Japan}
\affil[16]{Kavli Institute for Astrophysics and Space Research, Massachusetts Institute of Technology, 77 Massachusetts Avenue, Cambridge, MA 02139, USA}
\affil[17]{Lawrence Berkeley National Laboratory (LBNL), Berkeley, CA 94720, USA}
\affil[18]{University of California, Berkeley, Department of Physics, Berkeley, CA 94720, USA}
\affil[19]{Max-Planck Institute for Nuclear Physics, 69117 Heidelberg, Germany}
\affil[20]{Department of Physics, Chungnam National University, Daejeon 34134, Republic of Korea}
\affil[21]{Department of Physics, Massachusetts Institute of Technology, Cambridge, MA 02139, USA}
\affil[22]{Dipartimento di Fisica, Universit\`{a} di Torino, via P. Giuria 1, I--10125 Torino, Italy}
\affil[23]{Istituto Nazionale di Fisica Nucleare, Sezione di Torino, via P. Giuria 1, I--10125 Torino, Italy}
\affil[24]{Theoretical Physics Department, CERN, 1 Esplanade des Particules, CH-1211 Geneva 23, Switzerland}
\affil[25]{Department of Physics and Astronomy, University of Utah, Salt Lake City, UT 84112, USA}
\affil[26]{Department of Physics, Sungkyunkwan University, Suwon 16419, Korea}
\affil[27]{Department of Physics and Astronomy, University of Oklahoma, Norman, OK 73019, USA}

\begin{document}

\maketitle




\newpage

\section*{Executive Summary}
This whitepaper focuses on the astrophysical systematics which are encountered in dark matter searches. Oftentimes in indirect and also in direct dark matter searches, astrophysical systematics are a major limiting factor to sensitivity to dark matter. Just as there are many forms of dark matter searches, there are many forms of backgrounds. We attempt to cover the major systematics arising in dark matter searches using photons---radio and gamma rays---to cosmic rays, neutrinos and gravitational waves. Examples include astrophysical sources of cosmic messengers and their interactions which can mimic dark matter signatures. In turn, these depend on commensurate studies in understanding the cosmic environment---gas distributions, magnetic field configurations---as well as relevant nuclear astrophysics. We also cover the astrophysics governing celestial bodies and galaxies used to probe dark matter, from black holes to dwarf galaxies. Finally, we cover astrophysical backgrounds related to probing the dark matter distribution and kinematics, which impact a wide range of dark matter studies. In the future, the rise of multi-messenger astronomy, and novel analysis methods to exploit it for dark matter, will offer various strategic ways to continue to enhance our understanding of astrophysical backgrounds to deliver improved sensitivity to dark matter.

\newpage

\tableofcontents

\newpage 

\input{introduction.tex}

\input{synergies_indirect.tex}

\input{synergies_cosmology.tex}

\input{synergies_direct.tex}

\input{techniques.tex}

\input{conclusion.tex}


\bibliographystyle{unsrt}
\bibliography{main.bib}

\end{document}

%% file: introduction.tex
\section{Introduction}

Searches for dark matter (DM) are often driven by the ability of a particular experiment or set of experiments to observe the DM signal in its data. However, these searches are limited by our understanding of the experimental backgrounds and astrophysical systematic uncertainties surrounding the DM and its environment. \textit{In many ways, the search for signals from DM is the search to understand all emission which is not due to DM so the DM may shine through.} In this white paper, we will give an overview of the experimental and astrophysical backgrounds that can affect DM searches and the many ways that multiwavelength and multimessenger analysis can help identify, quantify, and remove these backgrounds to DM searches.

Below, we first provide an overview of three areas of DM search which can be improved through more careful understanding of the environments surrounding the DM - searches with cosmic messengers, searches in natural laboratories, and terrestrial searches for DM. With cosmic messengers, DM is being explored via "indirect detection"; here, DM is annihilating or decaying into Standard Model particles in an astrophysical environment which are then detected at Earth. Natural laboratories considers the extreme energies and distances accessible in astrophysical objects which can be altered due to the presence of DM. Finally, in terrestrial searches, DM "direct detection" is observed through DM interaction with matter on Earth. 

In all three areas of DM search covered here, there are large systematic uncertainties and backgrounds which need to be accounted for and reduced. In many of these cases, this can be addressed best through the use of multimessenger and multiwavelength astrophysical observations. Indeed, one experiment's background is another experiment's signal. Additionally, different techniques for probing astrophysics have different sources of uncertainty and different observables, which when combined can minimize experimental uncertainties. Tapping into the development of multiwavelength and multimessenger astrophysics provides powerful opportunities and solutions to address the needs of DM searches in the years to come.

\subsubsection*{Cosmic messengers}
The indirect detection of DM involves searches for Standard Model particles produced when DM annihilates or decays. These particles then travel to Earth where they are observed as an excess of experimental counts in the direction of the DM source. However, the interpretation of such observations as a DM signal is sensitive to a number of astrophysical assumptions. The morphology of the DM source, as well as its total DM density, can affect the interpreted DM cross-section or decay lifetime. The effects of particle propagation through the cosmos must be carefully understood to extract the energy spectrum and annihilation/decay final states of the DM at its source. For charged particles, it is also vital to understand propagation to extract the source positions from the particles' arrival directions. 

Each wavelength and messenger has different uncertainties and different strengths. Radio waves can observe synchrotron emission from pulsars but also from DM-produced $e^+/e^-$ pairs. Gamma-rays can show the most energetic astrophysical environments and DM annihilation spectra. Cosmic rays can probe particle transport in the galaxy as well as DM emission near the Earth. Neutrinos can be used to observe signals from distant high-energy sources and can probe the DM content of dense objects such as the Sun. Gravitational waves can provide information about black hole mergers but also black hole DM.

Many signatures from Standard Model particles and astrophysical processes can mimic DM behavior. Nuclei have transition lines that are similar to DM decay lines. Cosmic positrons and gamma-rays from pulsars can have similar energies to those from DM. Gamma rays from supernova remnants can cause excess counts in even the most DM-rich galaxies.  However, multiwavelength and multimessenger analyses are an ideal way to break these degeneracies in the data. A careful measurement of the interstellar magnetic field can improve both particle propagation concerns and reduce correlations between radio synchrotron observations and DM-produced $e^+/e^-$ pairs. Better understanding of energy spectral and spatial morphological information can distinguish DM versus astrophysical explanations of gamma-ray anomalies. Measurements of the diffuse gamma-ray flux can better quantify the gamma-ray signals which may cover up those from DM. Better quantification of nuclear abundances can show which line signals may be from DM. Joint experimental analysis and correlation between disparate datasets will be the hallmark of the next generation of DM searches.

\subsubsection*{Natural laboratories}
Similar to searches with cosmic messengers, natural laboratory searches for DM take place at astrophysical distances through interactions with Standard Model particles. However, instead of searching for DM annihilation or decay into those particles, these environments are directly modified due to the presence of DM. Due to the extreme environments encountered ---for example, extreme mass densities, temperatures, particle energies, just to name a few---astrophysical objects can be extremely sensitive to DM properties, more so than terrestrial experiments. However, the complication is the need to understand the astrophysical manifestation of Standard Model processes, which often lacks a full first-principles understanding due to the extreme scales involved. 
DM can cause the collapse of massive stars leading to modified nucleosynthesis or produce a previously unobserved class of quasars. On cosmic scales, DM can even affect the morphology of galaxies. Either in environments at extreme density and temperatures, or at cosmological distances and timescales, DM can shape its environment in ways that can provide unique insight into its nature. 

\subsubsection*{Terrestrial searches}
Direct detection searches for DM are often viewed as an alternative to astrophysical searches. Taking place on Earth, these experiments directly probe the nature of DM in a laboratory. However, even in direct detection experiments, the source of the DM is still astrophysical - the DM being probed is the DM halo surrounding the Earth and passing through it. And these DM have also affected and have been affected by astrophysical environments, as discussed above. More specifically, direct DM searches are sensitive to the velocity and spatial distribution of the DM in addition to its mass and nuclear coupling. 

Additionally, most direct detection DM searches have large backgrounds of astrophysical origin. Neutrinos, muons, and charged cosmic ray nuclei can pass through the Earth and mimic a DM signature. By correlating data from multimessenger and multiwavelength astrophysical experiments, these direct detection observatories can best minimize these backgrounds and search for DM interactions to even greater levels.

%% file: synergies_indirect.tex
\section{Synergies in indirect searches for dark matter with cosmic messengers}

Cosmic messengers are the primary method by which DM can be probed indirectly using astrophysical phenomena. Traditionally, photons and cosmic rays have been the messengers of choice largely due to their ease of detection, but in recent years they have been augmented by the rise of neutrino observatories and gravitational wave detections. In this section, we detail the important backgrounds encountered in when searching for DM with these cosmic messengers.

\subsection{Searches with photons}
\textit{Contributors: Esra Bulbul, Ilias Cholis, Shunsaku Horiuchi, Oscar Macias, Kohta Murase, Kerstin Perez, Marco Regis}\\

Among the cosmic messengers, photon detectors often deliver the most exquisite timing, angular, and energy resolutions. On the other hand, most if not all astrophysical objects/processes are prominent emitters of photons, which impact dark matter searches. In this section, we focus on two wavelengths---radio and gamma rays---which in particular hold significant promise for reducing astrophysical systematics in the future.

\subsubsection{Radio}

A variety of DM candidates are expected to show potentially detectable signatures at radio frequencies. 
The search technique and the systematic uncertainties related to the description of the astrophysical backgrounds strongly depend on the specific signature under investigation.
Here we outline three relevant cases.

\begin{itemize}
\item {\bf WIMPs and continuum synchrotron radiation:} 
DM particles with mass in the GeV-TeV regime can inject fluxes of primary and secondary high-energy electrons and positrons through annihilation or decay. 
Emitted into regions with ambient magnetic field at the $\mu$G level, such electrons and positrons give raise to a synchrotron radiation peaked in the radio band.
\textit{In order to characterize the $e^+/e^-$ equilibrium density, a good knowledge of the interstellar medium and of the turbulent and regular components of the magnetic field are needed.} 

The former allows to compute the energy losses associated to the cooling of $e^+/e^-$, while the latter is fundamental in the description of diffusion processes. Since the synchrotron signal comes from a convolution of the $e^+/e^-$ distribution with the magnetic field, the strength and spatial distribution of the latter are crucial ingredients.
The spatial distribution of the emission depends on the DM spatial profile, on the $e^+/e^-$ transport, and on the morphology of the magnetic field. 
The frequency spectrum is typically more curved than a power-law, but no striking spectral features are present.
Therefore, the disentanglement of the DM signal from an astrophysical emission might be rather awkward. This dictates the search to focus on regions with faint backgrounds~\cite{Regis:2014tga}, with the complication that those regions typically show low magnetic fields, whilst a significant magnetic strength is needed. The measurement of the magnetic field in dark structures (such as in dSph galaxies or in the external suburb of spiral galaxies and clusters) is a challenge, but a great advancement is expected in the coming years thanks to Faraday rotation measures of next-generation telescopes, such as the SKA~\cite{SKAMagnetismScienceWorkingGroup:2020xim}. 

\item {\bf Axion-like particles and spectral lines:} 
Axion-like particles (or dark photons) with mass around $\mu$eV can decay or convert into radio waves. The ALP-photon conversion (also called Primakoff effect) is particularly promising in objects having magnetic fields with significant strength and small coherence scale, e.g., magnetospheres of neutron stars~\cite{Sigl:2017sew,Hook:2018iia,Foster:2020pgt}; see also section \ref{sec:bodies}. The uncertainties on the theoretical prediction of the signal can be rather large and related to: a) the description of the magnetosphere (which requires both observations and simulations), b) the propagation of the radio photons generated from the conversion (a subject that is currently under quite active investigation in the community, see, e.g.,~\cite{Witte:2021arp}) and 3) the DM density in these objects.
In large-scale astrophysical environments, where the coherence scale of the magnetic field is larger, the rate of stimulated ALP decay into two photons~\cite{Caputo:2018vmy} supersedes that of ALP-photon conversion.
Here the uncertainties can be lower, as they are related to the description of the ambient photon field (responsible for the stimulated decay), which is reconstructed through the observed continuum flux of the source.
Also dark photon DM can oscillate resonantly into photons in thermal plasma, if the plasma frequency matches the dark photon mass. This might happen around stars, and the uncertainties in the model prediction are related to the description of the plasma properties and of the DM density in the stellar environments.

In all above cases, the signal is a spectral line with expected width $\sim \sigma_v/c\sim 10^{-2}-10^{-4}$, where the DM velocity dispersion $\sigma_v$ ranges from $\sim 10^3$ km/s in galaxy clusters to $\sim 10$ km/s in dSph. The experimental challenges involve the requirement of high frequency/angular resolution for the telescope, and the disentanglement of the DM-induced line from other radio lines that might be present in the field.

\item {\bf Ultra-light bosons and polarimetry:} 
The coupling between photons and ultra-light bosons can lead the left- and the right-circularly polarized light to travel at different velocities when crossing a DM halo (the birefringence
effect~\cite{Harari:1992ea}). This results in an achromatic rotation of the plane of polarization of linearly polarized light. Spectro-polarimetric observations at GHz-frequencies can be used to constrain DM candidates around the ``fuzzy" DM mass scale~\cite{Basu:2020gsy}. The key requirement is high sensitivity associated to rotation measures, as well as high spectral and angular resolutions.

\end{itemize}

\subsubsection{Gamma rays}

Gamma rays cover an ideal energy range for proving a range of DM candidates, including weak-scale DM and heavy DM. As such, substantial efforts have been devoted to search for continuous gamma rays, gamma-ray lines, and other more exotic signatures. It is crucial to construct improved astrophysical background models that allow a robust detection of these potential signals \cite{Fermi-LAT:2016afa}.
Table~\ref{tab:DMsignals_vs_BackgroundUncertainties} summarizes the  important background considerations: (i) the Galactic diffuse emission, due to interactions of energetic cosmic-ray particles with interstellar gas and radiation fields; (ii) unresolved/sub-threshold point sources, due to limited sensitivity of the gamma-ray facilities; and (iii) instrumental backgrounds. We show examples of DM search with gamma rays and which backgrounds are most relevant for them.

\begin{table}[ht!]
    \centering
    \begin{tabular}{|c|c|c|c|}\hline\hline
         & Galactic diffuse & Unresolved  & Instrumental \\
         & emission &  Sources  & Systematics \\\hline         
MeV DM              & $\bullet$ & $\bullet$ & \\\hline         
GeV DM @GC          & $\bullet$ & $\bullet$ & \\\hline
GeV DM @M31         & $\bullet$ & $\bullet$ & \\\hline
GeV DM @dSphs       &           &           & $\bullet$ \\\hline
GeV DM @clusters    &           & $\bullet$ & \\\hline
Heavy DM            &           &  $\bullet$         & $\bullet$ \\\hline
    \end{tabular}
    \caption{Summary of DM searches with gamma rays and the main categories of backgrounds. Bullet points illustrate the main relevant backgrounds; other backgrounds can still be relevant under specific conditions. Here, heavy DM refers to DM above the electroweak scale.}
    \label{tab:DMsignals_vs_BackgroundUncertainties}
\end{table}

\subsubsection*{Background systematics}

\noindent We first broadly cover the 3 categories of backgrounds and approaches to model them:
\begin{enumerate}
\item \textbf{Galactic diffuse emission:} is by far the most important and complicated background in DM searches with gamma rays. It arises from the interactions of energetic cosmic-ray particles with interstellar gas and radiation fields. In the literature, there are broadly two strategies to model it: (i) numerically solving cosmic-ray transport, e.g.,~\cite{Porter:2021tlr}; and (ii) data-driven phenomenological model, e.g.,~\cite{Fermi-LAT:2016zaq}. In both cases, \textit{the single most important challenge in generating accurate Galactic diffuse emission models remains obtaining accurate distributions of input ingredients, e.g., cosmic-ray sources, magnetic fields, and target gas densities.}

\item\textbf{Unresolved sources:} The cumulative emission of unresolved astrophysical gamma-ray sources could give rise to a significant diffuse background component which is challenging to model. Recent developments and future directions are discussed in Sec.~\ref{sec:novel_techniques}. 

\item\textbf{Instrumental systematics:} Typically, when searching for DM in crowded locations, such as the Galactic Center of the Milky Way galaxy, the dominant source of background is astrophysical. However, when pushing the sensitivity to DM down to the limits of detector capabilities, the issue of detector systematics grows. Examples of this include the use of gamma-ray quiet sources like dwarf galaxies, or using parts of sky with a lot less astrophysical gamma rays, e.g., high latitude regions. In turn, detector systematics varies widely depending on the gamma-ray energy range, detector technology, and also potentially target location. We do not attempt to cover the myriad of detectors and search situations here. Instead, in Sec.~\ref{section:synthesis}, we cover ideas for dealing with instrumental systematics in the context of combining or comparing results from multiple instruments. 

\end{enumerate}

Thus, improving the Galactic diffuse emission model is of primary importance for increasing the sensitivity of gamma rays to DM searches. Key progress in the future are:  

\begin{itemize}
    \item \textbf{Improved gas maps:} Three-dimensional maps of atomic hydrogen (H$I$) are needed. Current maps are limited by \textit{avoidable} simplifying assumptions such as idealized kinematics (assuming the gas follows circular orbits), and neglecting absorption features in the data~\cite{Macias:2016nev}. Biases introduced by these assumptions are some of the largest for extracting properties of the Fermi GeV excess (see accompanying dedicated whitepaper \cite{SnowmassCF1WP6}). One solution~\cite{Pohl:2007dz,Macias:2016nev} is to construct physically-motivated gas velocity profiles with hydrodynamic simulations, and solving the full radiative transfer equation (i.e., including 21-cm line emission, absorption, and continuous emission). This remains promising avenue for the future.
    Another direction is the development of new 3-D dust maps \cite{2019ApJ...887...93G} which may be of great benefit due to a known correlation between dust and gas. 
    
    \item \textbf{Improved gas models:} Molecular hydrogen cannot be measured directly. Instead, it is assumed to be well mixed with carbon monoxide (CO), which in turn can be traced by its 2.6 mm emission line. One of the necessary assumptions is that the H2 gas column density is proportional to the CO gas column density. However, studies~\cite{Burgh:2006bn,Liszt:2012uj} have shown that this molecular-hydrogen-to-CO conversion factor ($X_{CO}$) could vary significantly with distance. It is essential to find better tracers of Galactic H2 gas material and/or get an improved handle on the behaviour of the $X_{CO}$ in the Galactic environment. 
    In addition analyses that marginalize over all possible configurations of H2 maps need to continue being developed. 
    
    \item \textbf{Improved models for the interstellar radiation fields (ISRF):} the distribution of low-energy ambient photons (emitted by stars and re-emitted by dust) is a key ingredient in determining the spectrum and spatial morphology of the inverse Compton emission. State-of-the-art ISRF models include anisotropic features such as galactic arms and stellar bulge~\cite{Porter:2017vaa}. The inverse-Compton component shows strong degeneracy with the Fermi GeV excess~\cite{Fermi-LAT:2015sau} and thus further updates on the ISRF from multi-wavelength dataset will be important.

    \item \textbf{Improved models 
    and marginalization over Galactic magnetic fields:} Cosmic-ray electrons/positrons lose energy efficiently by gyrating in the Galactic magnetic fields. The uncertainty of current models for the random and regular magnetic fields in the Galaxy are large and may explain significant fractions of the Fermi GeV excess~\cite{Orlando:2013ysa}. It would be significant to advance the modeling of the total/polarized synchrotron radiation in order to separate these from other components. Obtaining better models for the magnetic fields will require global fits to the data in which diffuse gamma rays and radio (MHz$-$GHz) observations are considered.     
    
    \item \textbf{Beyond \textit{steady-state} solutions of the cosmic-ray transport equation:} The inverse-Compton maps generally used in the community are constructed under the assumption of \textit{steady-state} spatial distribution of sources and injection luminosities. However, the steady-state assumption breaks down for electrons/positrons with energies larger than approximately $100$ GeV because of their short cooling times~\cite{Porter:2021tlr}. 
    Dedicated modeling and data-analyses efforts will be required to obtain more realistic inverse Compton maps which agree with multi-messenger observations. 
    Also, the effects of anisotropic diffusion of cosmic rays associated with the large scale morphology of the galactic magnetic fields need to be accounted for. Anisotropic diffusion can have an impact also on the morphology of the inverse-Compton scattering maps \cite{Dobler:2011mk}.
    
    \item \textbf{Improved models for cosmic-ray sources:} this includes source types and their respective spectra as well as spatial distributions. Some sources are largely leptonic (e.g., pulsars) while others produce all species of cosmic rays (e.g., supernova remnants). 
    Also, these sources may enrich the ISM with distinctively different injection cosmic-ray spectra. Marginalizing over these uncertainties may be a solution going forward \cite{Calore:2014xka, Cholis:2021rpp}. 
    
    \item \textbf{High resolution simulations and multi-messenger fitting:} One solution to that is building up new physically motivated and high resolution galactic diffuse emission models. This is a complicated task requiring new modeling techniques, fits to new multi-wavelength data, and substantial computing resources. One critical improvement will be increasing the resolution of gas maps. Moreover, creating a multitude of galactic diffuse emission models for the inner galaxy allows to test a potential dark matter signal under a sequence of physical hypotheses and thus can alleviate any bias that may be introduced from any single one background emission model \cite{Calore:2014xka, Cholis:2021rpp}.

\end{itemize}

\subsubsection*{Diffuse quasi-isotropic gamma-ray background}

Searching for gamma rays in the diffuse gamma-ray background will enable us to probe heavy dark matter at GeV-TeV energies and above. 
\textit{However, astrophysical contributions often dominate the signal making their understanding crucial for making progress. }

The extragalactic gamma-ray background has been measured by the {\it Fermi} gamma-ray satellite~\cite{Fermi-LAT:2014ryh}, and about half of the sub-TeV gamma rays originates from blazars. The isotropic component of the diffuse component is so-called the isotropic diffuse gamma-ray background (IGRB), which includes contributions from unresolved point sources such as blazars~\cite{Ajello:2015mfa,TheFermi-LAT:2015ykq,Lisanti:2016jub}, cosmogenic gamma rays induced by cosmic rays, and any unaccounted-for Galactic emission. 
Thus, there is a rich mixture of backgrounds that need to be understood. For example: 

\begin{itemize}
    \item The IGRB has been used for constraints on annihilating DM~\cite{Srednicki:1985sf,Rudaz:1986db,Jungman:1994cg,Bergstrom:1997fj}. Depending on the clumping factor from DM substructures~\cite{Ng:2013xha}, a fraction of the IGRB could come from DM, and understanding astrophysical contributions to the IGRB as well as accurate modeling of the substructures are relevant for probing annihilating DM in the GeV-TeV range. 
    \item The IGRB measured by {\it Fermi} is powerful for heavy DM with masses above TeV energies~\cite{Ellis:1990nb,Gondolo:1991rn,Murase:2012xs,Cohen:2016uyg,Ishiwata:2019aet,Chianese:2021jke}. Extragalactic gamma rays initiate electromagnetic cascades in intergalactic space, and eventually contribute to the GeV-TeV IGRB. Cascade constraints are applicable to TeV or higher DM masses, up to even GUT scales with $10^{15}-10^{16}$~GeV~\cite{Murase:2012xs,Cohen:2016uyg}. 
    Knowledge of intergalactic photon fields and magnetic fields are crucial for accurate cascade calculations. 
    \item For decaying DM, both Galactic and extragalactic components contribute to the IGRB, and the Galactic contribution is crucial to constrain heavy DM with masses above TeV energies. Galactic gamma rays can reach Earth without strong attenuation via two-photon annihilation with the cosmic microwave background radiation, and they provide a powerful probe of heavy DM~\cite{Ahlers:2013xia,Murase:2015gea,Esmaili:2015xpa,Kalashev:2017ijd}. 
\end{itemize}

On the observation side, air-shower arrays, which are presumably cosmic-ray detectors such as KASCADE, have been used to place constraints on heavy DM in the sub-PeV range~\cite{Aglietta:1992aui,CASA-MIA:1997tns,KASCADEGrande:2017vwf,Dzhappuev:2020bkh}. Water Cherenkov detectors such as HAWC have also provided useful constraints in the 10-100~TeV range~\cite{HAWC:2017udy}. 
Recently, the Tibet AS$\gamma$ experiment reported the detection of diffuse Galactic gamma-ray emission~\cite{TibetASgamma:2021tpz}, and searches for high-latitude gamma-ray emission will be crucial for testing DM models for IceCube neutrinos (see section \ref{sec:neutrino}). LHAASO~\cite{He:2021rsg} will provide more stringent constraints on decaying DM.  

Finally, gamma-ray probes are intricately connected to cosmic-ray physics.  
Cosmic rays produced inside the Galactic plane may propagate into the halo region, but the detailed properties are uncertain. Revealing the roles of cosmic rays, including their feedback, in the halo region will help us to better model Galactic diffuse emission as well as to improve sensitivities to heavy DM. 

\subsection{Searches with cosmic rays}
\textit{Contributors: Ilias Cholis, Peter von Doetinchem, Kohta Murase, Volodymyr Takhistov}\\

In addition to time and energy information, cosmic rays contain rich composition information, including anti-particles. Observations of cosmic rays also reach substantially higher energies than other messenger particles, approaching $\sim 10^{20}$ eV limited only by their energy losses on the CMB. Thus, they remain a powerfully unique diagnostic of new physics. However, they also come with unique nuclear and astrophysical challenges, as described below. 

\subsubsection{Anti-matter cosmic rays}

Antimatter cosmic rays remain a major probe for the search of exotic physics in the Milky Way. Antiprotons and positrons are the only species observed by cosmic-ray detectors so far. Detailed models and simulations for the possible astrophysical sources of such species as close-by pulsars and supernova remnants need to be developed that envelope the uncertainties on the physical properties of these sources; which includes their distribution in space, their power output including its time-evolution, the type of particles they inject into the interstellar medium and their respective spectra. Also models need to include the uncertainties of cosmic-ray propagation through the interstellar medium and the heliosphere \cite{Cholis:2017qlb, Hooper:2017gtd, Hooper:2017tkg, Cholis:2018izy, Mertsch:2018bqd, Evoli:2020szd, Evoli:2021ugn, Cholis:2021kqk}. As such sources appear in a stochastic manner in space and time only simulations that account for that and probe their properties as an averaged population can be reliably developed as e.g. \cite{Mertsch:2018bqd,Cholis:2021kqk}.  

In addition, a recent excess in $\sim  10$ GeV energy antiprotons observed by \textit{AMS-02} has been claimed \cite{Cuoco:2016eej, Cui:2016ppb}. That excess is quite robust to the astrophysical uncertainties of local cosmic-ray propagation and cross-sectional uncertainties and suggest a signal of dark matter with similar properties to those required to explain the galactic center excess \cite{Cholis:2019ejx, Cuoco:2019kuu, Hooper:2019xss}. However, a full scrutiny on the systematic errors of the \textit{AMS-02} instrument remains a necessity \cite{Boudaud:2019efq, Heisig:2020nse}. 
Further discussions between the experimental and theoretical communities, in order to achieve a better modeling of the correlation matrix of the instrumental systematic \textit{AMS-02} errors, would allow us to settle on whether the antiproton excess is of instrumental or of astrophysical origin.
Of those errors likely the most significant in the $\sim  10$ GeV energy range of the antiproton excess are the effective acceptance and the nucleon-nucleon cross-section vs energy assumptions.  

Finally, the possible detection of antideuteron or anti-helium cosmic-rays by \textit{AMS-02} or GAPS will have a profound impact on our ability to probe dark matter in the Milky Way as the astrophysical backgrounds for such cosmic rays are strongly suppressed \cite{Korsmeier:2017xzj, Poulin:2018wzu, Cholis:2020twh, GAPS:2020axg}.
If GAPS with its expected $\sim 3$ month flight is successful in providing a sensitivity at lower antideuteron energies to that of \textit{AMS-02}, follow-up observations will be of great value. As GAPS will be sensitive at $O(0.1)$ GeV in kinetic energy per nucleon, cosmic-ray propagation uncertainties and especially those through the heliosphere responsible for the time-dependent solar modulation of cosmic rays, will need to be further reduced. To that end the \textit{AMS-02} time-dependent measurements \cite{AMS:2018qxq, AMS:2018avs} can be of great importance \cite{Zhu:2018jbk,Tomassetti:2019tyc, Cholis:2020tpi}.  
Moreover, significant reduction on the production cross-section uncertainties for these species will be of paramount importance \cite{vonDoetinchem:2020vbj}. To that end, a better modeling of the forward production of cosmic rays in the collision reference frame and a better understanding of the coalescence model implemented and possible corrections to it will be important. Further measurements from fixed target accelerator experiments in a range of energies with the capacity to detect antiproton, antideuteron and anti-helium nuclei (such as those presented in \cite{Kappl:2014hha,Graziani:2017als}) will be of great benefit to achieve those reduced cross-sectional uncertainties. Examples of such experiments are discussed in Refs.~\cite{na61, Adams:2018pwt, ALICE:2014sbx}. A more detailed discussion of these cross-sections can be found in Section 4 of the white paper "Snowmass2021 Cosmic Frontier: The landscape of cosmic-ray and high-energy photon probes of particle dark matter" \cite{SnowmassCF1WP5}.

Cosmic rays interacting along our line of sight could serve as sensitive astrophysical messengers~\cite{Essey:2010er,Murase:2011cy,Takami:2013gfa,Kochocki:2020iie}.
Cosmic rays directly colliding with DM can allow probing novel parameter space compared to conventional searches~\cite{Cappiello:2018hsu,Bringmann:2018cvk,Ema:2018bih,Jho:2021rmn,Bell:2021xff}. If DM carries an electric charge it could be accelerated in astrophysical environments such as supernovae remnants akin to standard cosmic rays~\cite{Chuzhoy:2008zy,Li:2020wyl}, providing ``dark cosmic ray''~\cite{Hu:2016xas} messengers that can be effectively studied in neutrino laboratories.

\subsubsection{Atmospheric collider}

Collisions of energetic cosmic rays provide a natural laboratory for exploration of DM and new physics. As isotropic cosmic ray flux collides with the atmosphere, in analogy with conventional colliders, copious amounts of particles are produced. The ``atmospheric collider'' has been historically employed as a central tool for analyzing neutrinos, leading to discovery of neutrino oscillations~\cite{Super-Kamiokande:1998kpq}. Recently, atmospheric collider has been identified as a unique tool for novel exploration of light dark matter~\cite{Alvey:2019zaa}, milli-charge (DM) particles~\cite{Plestid:2020kdm,ArguellesDelgado:2021lek} and magnetic monopoles~\cite{Iguro:2021xsu}. Unlike conventional colliders, atmospheric collider is always on and provides a robust universal flux of particles for all terrestrial experiments. This allows to make direct comparisons, challenging to do otherwise, of ambient flux searches with e.g. collider searches. Further explorations as well as more detailed simulation studies are needed to explore fully capabilities and complementary with other searches of atmospheric collider.

\subsubsection{Ultrahigh energies}

The origin of ultrahigh-energy cosmic rays (UHECRs)---cosmic rays reaching $\sim 10^{20}$ eV---remains unknown. 
However, the existence of a spectral cutoff around $50-60$~EeV~\cite{HiRes:2007lra,PierreAuger:2008rol,PierreAuger:2020kuy} strongly supports their astrophysical origin; top-down scenarios~\cite{Aloisio:2006yi}, including super-heavy DM models~\cite{Chung:1997rz,Berezinsky:1997sb}, for the dominant origin of UHECRs are very unlikely.   
Nevertheless, this does not spoil the possibility that DM is super-heavy. Super-heavy DM, whose masses may be in the GUT scale, can be produced via various scenarios, e.g., gravitational production in the early universe~\cite{Chung:1998zb,Kuzmin:1998uv}. It is possible that such heavy DM is meta-stable, and the DM could give a sub-dominant contribution to the UHECR flux~\cite{Aloisio:2007bh}.  
In the decaying DM scenario, the Galactic contribution is important~\cite{Dubovsky:1998pu}, and protons and neutrons from the Galactic halo are not depleted. The UHECR data, especially above the cutoff, provide constraints on super-heavy DM, which are complementary to neutrino and gamma-ray limits. 

In the future, revealing UHECR sources will enable us to explore large parameter spaces through spectral and anisotropy observations. It is important to also better understand intergalactic magnetic fields, which affects UHECR propagation, as well as uncertainties in photonuclear reactions, which impacts UHECR energy loss and composition changes, and questions such as the transition from Galactic to extragalactic flux components. Also, since UHECRs from decaying DM are dominated by protons and neutrons, better understanding of the UHECR composition will be helpful.

Future probes also stand to benefit from multi-messenger astronomy. For example, constraints on decaying DM depend on the final states which are unknown. If the DM decays into Standard Model particles, only a fraction of them can be nucleons, while a large fraction of the parent DM energy can be carried by ultrahigh-energy gamma rays and neutrinos~\cite{Birkel:1998nx,Sarkar:2001se,Barbot:2002ep}. UHECR observations are also sensitive to ultrahigh-energy gamma-rays and neutrinos, providing another probe of super-heavy DM~\cite{Kalashev:2020hqc}.

\subsection{Searches with neutrinos}\label{sec:neutrino}
\textit{Contributors: Shunsaku Horiuchi, Kohta Murase, Jong-Chul Park, Carsten Rott, Volodymyr Takhistov, Yun-Tse Tsai}\\

Depending on the scenario, DM can manifest in neutrinos over a broad spectrum of energies, making neutrino telescopes excellent instruments for indirect DM detection. The Super-Kamiokande large underground water Cherenkov experiment provides fruitful detection prospects for DM-associated neutrino signatures from above 30 MeV to well over TeV, coming from WIMPs with masses of $\sim$GeV to $\sim$TeV.
Characteristic WIMP annihilation channels include $b\overline{b}$, $W^+W^-$, $\mu^+\mu^-$ and $\nu\overline{\nu}$.
Stringent limits have been placed by Super-Kamiokande and IceCube on signatures associated with annihilation of DM in the Sun~\cite{Super-Kamiokande:2015xms,Aartsen:2016zhm} as well as Galactic Center~\cite{Super-Kamiokande:2020sgt,Aartsen:2017ulx,Abbasi:2011eq}. 
Atmospheric neutrinos constitute a significant source of background for these searches \cite{Beacom:2006tt,Yuksel:2007ac}. While this is generally well-known around GeV--TeV energies, there remains systematic uncertainties arising from interaction cross-sections and complicated channels at higher and lower energies. In the next decade solar DM searches will also face a background of atmospheric neutrinos in the solar atmosphere, resulting in a solar atmospheric neutrino floor~\cite{Arguelles:2017eao,Ng:2017aur,Edsjo:2017kjk,Masip:2017gvw}. Energy resolution would be the key to distinguish these from a DM signal by exploiting the  differences in energy spectra. On the other hand, searches for solar DM are extremely robust against astrophysical uncertainties in the underlying DM velocity distribution~\cite{Danninger:2014xza,Choi:2013eda,Nunez-Castineyra:2019odi}. 

Aside from traditional searches for DM annihilations, neutrino telescopes can be readily exploited to explore other DM scenarios. For example, Super-Kamiokande has set leading limits on boosted DM originating from the Sun and Galactic Center~\cite{Agashe:2014yua, Berger:2014sqa, Kong:2014mia, Kim:2016zjx, Super-Kamiokande:2017dch} as well as ``dark cosmic rays'' of DM carrying electric charge and accelerated in astrophysical sources~\cite{Hu:2016xas}.
In addition, the Deep Underground Neutrino Experiment (DUNE), the next-generation neutrino detector based on the technology of liquid-argon time-projection chamber (LArTPC), is evaluating its sensitivity to boosted DM through electron and/or hadronic scatterings~\cite{Necib:2016aez, Alhazmi:2016qcs, Kim:2020ipj, DeRoeck:2020ntj, DUNE:2020fgq, PhysRevD.103.095012}.
Although neutrino detectors have higher detection thresholds than conventional direct DM detectors, neutrino detectors are more massive and detectors based on different technologies can probe complementary parameter space~\cite{Kim:2020ipj}.

Since boosted DM and their signatures span a wide energy range, the important backgrounds also vary. While atmospheric and astrophysical neutrinos are the main background in the energy range greater than sub-GeV, solar neutrinos and radiological background have significant impacts on MeV-scale signals.  Knowledge of neutrino fluxes from natural sources and of aforementioned neutrino interaction cross sections will enhance the sensitivity on boosted DM searches.  In particular, sophisticated analysis techniques require reconstructed kinematics to discriminate signal from background, and understanding kinematics of neutrino interactions via related cross section measurements will be beneficial.  Further, boosted DM searches through hadronic scattering often look for signatures similar to neutral-current neutrino interactions, and an efficient muon-pion separation will help reduce the contribution from current-current events.  In the MeV regime, the radiological background makes the triggering system challenging, and to cope with this situation, a number of ideas inspired by conventional DM search experiments, such as deploying underground argon in LArTPC-based detectors to reduce Ar 39 isotopes, are being discussed.

The observed IceCube neutrinos are most likely astrophysical, which is supported by the reported coincidences with some astrophysical sources~\cite{IceCube:2019cia}. Nevertheless, heavy DM could contribute to the diffuse neutrino flux, and various DM models have been discussed in light of the IceCube data. 
Because of the unitarity bound, many scenarios rely on DM decay into SM particles~\cite{Feldstein:2013kka,Esmaili:2013gha,Bhattacharya:2014vwa,Fong:2014bsa,Higaki:2014dwa,Rott:2014kfa,Murase:2015gea,Dev:2016qbd,DiBari:2016guw,Kachelriess:2018rty}. 
The diffuse neutrino flux is measured over a wide energy range from 10~TeV to PeV energies, in which decaying DM scenarios involving quarks in the final state are in tension with the IGRB data~\cite{Murase:2015gea}, especially for models explaining the 10-100~TeV neutrino flux. 
Neutrinophilic DM or DM decaying into non-SM particles may evade the present gamma-ray constraints with a partial contribution to the IceCube neutrino flux~\cite{Feldstein:2013kka,Higaki:2014dwa,Boucenna:2015tra,Chianese:2018ijk,Hiroshima:2017hmy}, and signatures of neutrino-DM scattering are also considered~\cite{Arguelles:2017atb,Murase:2019xqi}.     
The IceCube experiment also has provided stringent limits on neutrino signatures associated with annihilation~\cite{ANTARES:2020leh} or decay~\cite{IceCube:2018tkk} of DM.

Even if heavy DM is not responsible for the bulk of the diffuse neutrino flux, neutrino searches with IceCube/KM3Net/Baikal-GVD are powerful for DM with mass ranges above TeV energies. Neutrinos can reach Earth even at ultrahigh energies, where the atmospheric neutrino background is negligibly small and the astrophysical background from cosmogenic neutrinos may also be low depending on the UHECR composition. 
Ultrahigh-energy neutrinos serve as a powerful probe of heavy DM even at the GUT or Planck scale, and their searches with various neutrino and UHECR detectors such as Auger are complementary to gamma-ray and cosmic-ray observations~\cite{Chianese:2021htv}. 

Better understanding of the origin of the diffuse neutrino flux will enable us to further explore the parameter space of heavy DM. Resolving the sources of IceCube neutrinos and improving the modeling of astrophysical components can be used to improve the constraints on the annihilating cross section and decay lifetime. In either DM annihilation or decay, DM models predict anisotropy following the Galactic DM distribution~\cite{Bai:2013nga}, which can be tested with more statistics achieved by upcoming neutrino observations. 

Individual source searches are also useful, and local DM halos such as galaxy clusters (e.g., Virgo) and galaxies (e.g., M31) will provide complementary tests for the DM contribution to the diffuse neutrino flux~\cite{Murase:2015gea}. This will be discussed later, and it has great synergies with neutrino frontier white paper on ``natural source''.

\subsection{Searches with gravitational waves}
\textit{Contributors: Ilias Cholis, Kuver Sinha, Volodymyr Takhistov}\\

Gravitational waves (GWs) can be used to search for primordial black holes (PBHs) with masses between $\sim 1$ and several 100 solar masses \cite{Bird:2016dcv, Sasaki:2016jop, Carr:2017jsz}. The current and upcoming GW observatories, can be used to search for and constrain the stellar mass range of that wide mass spectrum. 
In the mass ranges of the current and upcoming GW observatories, PBHs could contribute a sizable fraction of DM energy density $f = \Omega_{\rm PBH}/\Omega_{\rm DM}$~\cite{Macho:2000nvd,Ali-Haimoud:2016mbv,Poulin:2017bwe,Lu:2020bmd,Tsai:2020hpi,Takhistov:2021aqx}, with GW data suggesting $f \lesssim \mathcal{O}(10^{-3})$~\cite{Bird:2016dcv,Hutsi:2020sol,DeLuca:2021wjr,Sasaki:2016jop}, although uncertainties remain \cite{Jedamzik:2020omx} and need to be explored further. Variety of handles can be used to discriminate PBHs with astrophysical BHs (ABHs) via GW observations:
\begin{itemize}
    \item The mass spectrum of the detected GW events. With upcoming LIGO / VIRGO / KAGRA observations the BH mass spectrum will be measured with unprecedented precision. PBHs may cause a break, a bump or multiple bumps in that mass spectrum \cite{Kovetz:2017rvv, LIGOScientific:2020kqk}. 
    \item The spin of PBHs, which could be significant~\cite{Cotner:2018vug,Cotner:2019ykd,Harada:2017fjm}, and how that evolves with time~\cite{DeLuca:2020bjf} and for different dark matter halo masses. Spinning PBHs will also be distinguishable by their associated emission signatures such as jets~\cite{Takhistov:2021upb}.
    \item Merger rates and formation of PBH vs.~ABH binaries. PBH-PBH binaries can receive significant contributions from the late Universe~\cite{Bird:2016dcv} as well as early Universe before matter-radiation equality~\cite{Sasaki:2016jop}. In contrast, PBH-NS binaries follow stellar evolution and can only form at low redshifts. As demonstrated in~\cite{Sasaki:2021iuc}, PBH-NS merger rates are subdominant to the astrophysical BH-NS merger rates, suggesting that such identified events are of astrophysical origin even if PBH-PBH mergers significantly contribute to the GW data. 
    \item Recently observed merger events by LIGO/Virgo, including GW170817 \cite{LIGOScientific:2017vwq}, GW190425 \cite{LIGOScientific:2020aai}, GW190521 \cite{LIGOScientific:2020iuh}, and GW190814 \cite{LIGOScientific:2020zkf}, were discussed to be potential NS-PBH \cite{Tsai:2020hpi} or BH-PBH/PBH-PBH merger events \cite{Clesse:2020ghq}. NSs and PBHs of similar mass, or transmuted BHs from DM consuming NSs, are difficult to distinguish individually with GWs~\cite{Takhistov:2017bpt}. Statistical test allows to establish the origin of such events~\cite{Takhistov:2020vxs}. The excellent localization of GW170817 from the multi-messenger measurements provide detailed environmental information to test such hypothesis \cite{Tsai:2020hpi}.
    \item Searching for events with non-zero eccentricities. Such non-zero eccentricity events are expected from merging of PBH binaries \cite{Cholis:2016kqi, Clesse:2016ajp}. However, BH binaries in dense stellar environments as globular clusters or nuclear stellar clusters and at the environments around the central super-massive black holes of galaxies may also contribute \cite{Samsing:2017xmd, Gondan:2017wzd}. 
    \item The GW stochastic background at LIGO and future GW observatories. PBH binaries existed and merged at far greater cosmological redshifts than regular stellar mass BHs and can contribute to the GW stochastic background at frequencies where it is suppressed \cite{Mandic:2016lcn}. Even more so, the mechanism(s) responsible for the formation of PBHs may give a significant contribution to the GW background to be measured by LISA and pulsar timing arrays \cite{Nakama:2016gzw, Bartolo:2016ami}.   
\end{itemize}

To explore the above, a deeper understanding of astrophysical backgrounds i.e. the merger rates and properties of binaries composed of ABHs is required. In particular, the environments where ABH binaries exist, binary formation channels as well as evolution to their detected mergers needs to be extensively modeled.  

Beyond GWs, stellar and intermediate-mass BHs can be naturally accompanied by radiation emission associated with accretion disk formation~\cite{Lu:2020bmd,Takhistov:2021aqx} as well as outflows such as jets~\cite{Takhistov:2021upb}. This can result in variety of observable signatures, including heating of interstellar medium gas~\cite{Lu:2020bmd,Takhistov:2021aqx,Takhistov:2021upb}, which can be used to explore and constrain PBHs over a wide mass-range.

\subsection{Systematics due to dark matter distribution}
\textit{Contributors: Shin'ichiro Ando, Ilias Cholis, Lina Necib}\\

Indirect detection depends on the density of DM at the location in which DM is expected to decay or annihilate. Such locations are the Galactic Center, dwarf satellite galaxies, nearby galaxies such as Andromeda, or galaxy clusters. However, correctly estimating the DM density profiles is dominated by uncertainties, which we tackle below for the targets that can give the most competitive limits. 
\begin{itemize}
    \item \textbf{Galactic center}
    The galactic center or inner galaxy due to its great astrophysical complexity is oftentimes analyzed by imposing masks that extract from the analysis of point sources and dense interstellar medium gas regions. Such masks remove from the analysis also regions where the dark matter distribution may be the highest. Regions at latitudes above the Galactic disk may be preferable in searching for a dark matter signal. Such regions can still be sensitive in probing the dark matter signal and its distribution profile \cite{Daylan:2014rsa, Calore:2014xka, Cholis:2021rpp}. However, to robustly probe the inner degrees of the Galaxy the use of novel techniques in masking point sources or modeling them independently will be necessary. 
    \item \textbf{Dwarf galaxies density profiles} 
    Estimating the density of DM in dwarf galaxies challenging, as the only information available is the line-of-sight measurements of the velocities of member stars. One of the most commonly used methods to extract the DM density profile from the stellar line-of-sight velocities is Jeans analysis \cite{1915MNRAS..76...70J}, where assumptions about equilibrium, sphericity, and isotropy of these systems have to be made. \cite{Chang:2020rem,Guerra:2021ppq} present these issues using idealized mocks, where even in spherical, isotropic systems in equilibrium, a large number of stellar measurements (order of 10,000 stars) is required to robustly disentangle a core from a cusp. Alternatively, mass estimators methods \cite{Wolf:2009tu,Walker:2011zu}, rotation curves \cite{Rhee:2003vw,Valenzuela:2005dh}, the use of higher order moments of the velocities, the possible inclusion of proper motions when available \cite{Strigari:2007vn,Lazar:2019hkz,Guerra:2021ppq}, and non-spherical mass modeling \cite{Hayashi:2020jze} are being pursued to reduce the systematic uncertainties of the DM density profiles, making it an active area of research. It is also pointed out that adopting realistic priors will have a large impact on the estimate of density profiles for the ultrafaint dwarf galaxies \cite{Ando:2020yyk}.
\end{itemize}

%% file: synergies_cosmology.tex
\section{Synergies in indirect searches for dark matter with natural laboratories}

Nature provides extreme environments which are far beyond the capabilities of terrestrial detectors, and these can be advantageously used for DM search. For example, the core of a core-collapse supernova reach nuclear densities and MeV thermal temperatures; compact objects have magnetic fields reaching $\sim 10^{16}$ G; galaxies contain billions of stars and huge gravitational potentials; and natural accelerators generate cosmic rays at least up to $\sim 10^{20}$ eV in energy. In this section, we cover the astrophysical processes which become important backgrounds or limiting factors in the use of astrophysical phenomena as natural laboratories for DM search. 

\subsection{Searches with celestial bodies}\label{sec:bodies}
\textit{Contributors: Rebecca Leane, Carsten Rott, Volodymyr Takhistov}\\

Astrophysical compact objects can serve as natural laboratories for exploring dark matter. A variety of signals are possible, depending on the underlying dark matter model and the particular celestial object. DM in the Galactic halo can be captured by celestial bodies, and alter their properties. Particle DM can accumulate, and annihilate to either increase the celestial body's temperature~\cite{Bertone:2007ae,Freese:2008hb,Baryakhtar:2017dbj,NSvIR:Pasta,Leane:2020wob}, or produce detectable Standard Model products at its surface, such as gamma rays and neutrinos~\cite{Batell:2009zp, Meade:2009mu, Schuster:2009fc, Schuster:2009au, Bell:2011sn,Leane:2017vag,HAWC:2018szf,Nisa:2019mpb,Leane:2021tjj,Leane:2021ihh}. If DM is non-annihilating it can accumulate in large abundance, or if it is a primordial black hole in the asteroid $\sim 10^{-17}-10^{-12} M_{\odot}$ mass-range, the host can be destroyed, resulting in a variety of astrophysical signatures~\cite{Capela:2013yf,Fuller:2017uyd,Takhistov:2017bpt,Bramante:2017ulk,Takhistov:2017nmt,Kouvaris:2018wnh,Genolini:2020ejw,Takhistov:2020vxs,Dasgupta:2020mqg,Tsai:2020hpi}. 

Stellar destruction signatures include the potential production of $r$-process heavy element nucleosynthesis material~\cite{Takhistov:2017bpt,Bramante:2017ulk}, fast radio bursts~\cite{Fuller:2017uyd,Abramowicz:2017zbp,Kainulainen:2021rbg}, 511 keV radiation~\cite{Fuller:2017uyd}, formation of a new class of microqusars~\cite{Takhistov:2017nmt} as well as ``orphan'' gamma-ray bursts~\cite{Takhistov:2017nmt} and kilonovae~\cite{Fuller:2017uyd,Bramante:2017ulk} not accompanied by strong gravitational wave emissions. Intriguingly, such asteroid-mass PBHs are unconstrained and can constitute the entirety of the dark matter abundance. After NS implosions, there will remain $\lesssim 2.5 M_{\odot}$ ``transmuted''~\cite{Takhistov:2017bpt} solar-mass black holes, which are not expected from conventional stellar evolution and hence constitute a promising target for new physics searches~\cite{Takhistov:2017bpt,Bramante:2017ulk,Kouvaris:2018wnh,Takhistov:2020vxs,Dasgupta:2020mqg,Tsai:2020hpi}. The origin of detected solar-mass BH candidate events could be identified using statistical method based on BH mass-function proposed in~\cite{Takhistov:2020vxs}. These solar-mass BH (distinct from the BH with primordial origins) can also merge with other compact objects, e.g. BH or NS , providing GW and multi-messenger signatures \cite{Takhistov:2017bpt,Tsai:2020hpi}.
Detailed studies of stellar evolution,formation of astrophysical black holes, emission and nucleosynthesis associated with compact objects as well as reinvigorated observational campaigns for distinct and multi-messenger signatures are essential to further explore the nature of dark matter and its interactions with celestial bodies.

\subsection{Searches with galaxies}
\textit{Contributors: Mike Boylan-Kolchin}\\[0.3cm]

\indent As the systems that provided much of the original astrophysical motivation for taking dark matter seriously, galaxies are one of the most natural indirect probes of the nature of dark matter \cite{faber1979}. In the prevailing cold dark matter (CDM) model, virialized dark matter halos can form on scales that are much lower in mass than the stellar mass of the faintest galaxies: in a traditional WIMP picture, the lowest-mass dark matter halos are expected to be comparable in mass to the Earth ($10^{-6}\,M_{\odot}$; \cite{green2004}), and each of these dark matter halos should --- in the absence of the effects of galaxy formation --- possess a density distribution of $\rho \propto r^{-1}$ near its center \cite{navarro1997}, where the galaxy resides. These predictions of CDM, abundant substructure and central density cusps (as opposed to density cores, with no radial variation of the density profile near the center), have been the source of tension between the CDM model and observations \cite{bullock2017}. These tensions include:
\begin{itemize}
    \item the \textit{missing satellites problem} \cite{klypin1999, moore1999}: given the abundance of predicted substructure, why do we observe a relative paucity of satellite galaxies around the Milky Way?
    \item the \textit{cusp-core problem} \cite{flores1994}: contrary to the predicted central cusps found for galaxy-scale halos in $N$-body simulations of CDM structure formation, many galaxies exhibit evidence of density cores
    \item the \textit{too big to fail problem} \cite{boylan-kolchin2011a, boylan-kolchin2012}: a standard solution to the missing satellites problem is to posit that satellite galaxies can only form in the most massive subhalos of the Milky Way, with lower mass structure unable to collect or efficiently cool the gas required to form stars. In this scenario, the central masses of the predicted population of galaxy-hosting subhalos significantly exceed the measured masses of the observed satellite galaxies
    \item the \textit{diversity problem} \cite{oman2015, kaplinghat2020}: at a fixed galaxy rotation velocity, which is a proxy for the mass of the host dark matter halo, galaxies exhibit a wide range of behaviors of the inner rotation curve, from a quick rise near the center to a more slow and gradual increase. Why do galaxies formed in gravitational potential wells of similar depth have very different properties?
\end{itemize}

These ``small-scale challenges" to the $\Lambda$CDM model \cite{bullock2017, del-popolo2017} have provided motivation for detailed study of alternate dark matter models, including:
\begin{itemize}
    \item warm dark matter (WDM; \cite{bond1982, bode2001}), where dark matter has a non-negligible free-streaming length in the early Universe (before matter-radiation equality). Some sterile neutrino models fall in this category.
    \item self-interacting dark matter (SIDM), where dark matter has a non-negligible self-scattering cross section, affecting the distribution of dark matter in high-density regions.
    \item fuzzy dark matter (FDM; \cite{hu2000, peebles2000}), where dark matter has a de Broglie wavelength comparable to the sizes of the smallest galaxies, meaning that quantum pressure sets a minimum mass scale of collapsed objects.
    \item primordial black hole (PBH; \cite{carr1974, bird2016}) dark matter, where dark matter is composed of black holes with masses of $\mathcal{O}(10\,M_{\odot})$. Unlike CDM, gravitational interactions between individual dark matter ``particles'' and stars can then be important.
\end{itemize}
The effects of these non-CDM models of dark matter on properties of galaxies can be grouped into two broad classes: 
(i) the {abundance} of galaxies --- particularly low-mass galaxies --- is sensitive to the linear matter power spectrum (e.g., \cite{schneider2012}), and (ii) the {internal structure} of galaxies across a range of scales can further constrain non-gravitational dark matter interactions with itself or with baryons (e.g., \cite{bullock2017}). \textit{However, both of these probes --- the abundance and internal structure of galaxies --- are subject to potentially large uncertainties that originate from the galaxy formation process itself and gravitational coupling of dark and baryonic matter \citep{governato2012, brooks2014, onorbe2015, brooks2017, zavala2019}.}

A natural prediction of any dark matter model that affects the linear matter power spectrum (via, e.g., free-streaming or quantum pressure) is a truncation in the luminosity function of galaxies, either locally or in the high-redshift Universe \cite{maccio2010a, schneider2012, lovell2012}. However, the interpretation of a cut-off in the number of galaxies below a given mass scale is complicated by the effects of galaxy formation and observational limits. Cosmic reionization is widely expected to set a ``floor" to galaxy formation by suppressing the accretion of gas onto low-mass halos after a redshift of ${\sim}10$ \cite{bullock2000, somerville2015, rodriguez-wimberly2019}. Galaxy counts in the Local Group are therefore sensitive not just to the abundance of low-mass dark matter halos but also to the physics of galaxy formation in such objects. At high redshifts, the ultraviolet (UV) luminosity function of galaxies should provide a means of probing low-mass dark matter halos, and the appearance of a cut-off in the UV luminosity function could be indicative of dark matter physics that reduces the amplitude of the matter power spectrum on a corresponding scale \cite{menci2017}. However, the episodic nature of high-redshift star formation combined with the existence of galaxies at the present day that \textit{must} have been fainter than the detection limits of current or forthcoming telescopes \cite{weisz2017} complicates interpretations.

Virialized dark matter halos under the influence of gravity alone have been extensively studied and quantified using numerical simulations; such systems are known to be well-approximated by Navarro-Frenk-White density profiles \cite{navarro1997}, with $r^{-1}$ cusps on small scales and a $r^{-3}$ fall-off near the virial radius. Introducing appreciable non-gravitational interactions can alter these predictions, resulting in density cores (from, e.g., self-scattering or quantum pressure). However, the effects of galaxy formation and assembly can also result in a modification of the structure of dark matter halos. These effects can even go in both directions (increasing and decreasing dark matter densities). Some of the main relevant processes are the infall of baryonic matter, which can increase the dark matter density on galactic scales under certain circumstances \cite{blumenthal1986, gnedin2004}, and energy and momentum input from stellar evolution and black holes, which typically reduce the both the amount of dark matter and the slope of the dark matter density profile in galaxies' centers \cite{di-cintio2014, onorbe2015, brooks2017, tollet2016, fitts2017, dashyan2018}. 

Assumptions when going from modeling the distribution of luminous matter to inferring the distribution of dark matter are also a source of uncertainty. For example, non-circular motions and inclination variations in galactic disks affect inferences regarding the underlying total (dark matter plus baryonic) gravitational potential \cite{oman2019}, and stellar kinematic data are often limited to line-of-sight velocities, leaving the velocity anisotropy of stellar orbits as a potentially large systematic when converting from velocities to the underlying mass distribution \cite{wolf2010}. Even the stellar initial mass function (IMF) plays a role in modeling and contributes to the uncertainty budget: the stellar mass contributed by an observed population of stars depends on the stellar mass-to-light ratio, which is set by the IMF \cite{thomas2011}. This is a particularly important issue for gravitational lensing studies of dark matter.

%% file: synergies_direct.tex
\section{Synergies in direct detection of dark matter}

\subsection{Astrophysical backgrounds}
\textit{Contributors: Sebastian Baum, Shunsaku Horiuchi, Ibles Olcina, Volodymyr Takhistov}\\

Solar, supernova and atmospheric neutrinos constitute irreducible astrophysical backgrounds for direct DM detection experiments. Such ``neutrino floor'' \cite{Billard:2013qya,Monroe:2007xp,Ruppin:2014bra, Gelmini:2018ogy} (or alternatively defined ``neutrino fog'' \cite{OHare:2021utq}) could significantly reduce sensitivity of future DM searches, especially in the region where the recoil spectrum of the neutrino background is approximately degenerate with DM signatures. 
Hence, the primary questions are: {\it what is the future progress in characterizing astrophysical neutrino fluxes, and given these improvements, what can we learn about the nature of the dark matter and BSM physics using signals below the present bounds from direct detection experiments?}

\begin{figure*}
\begin{center}
\includegraphics[width=0.65\textwidth]{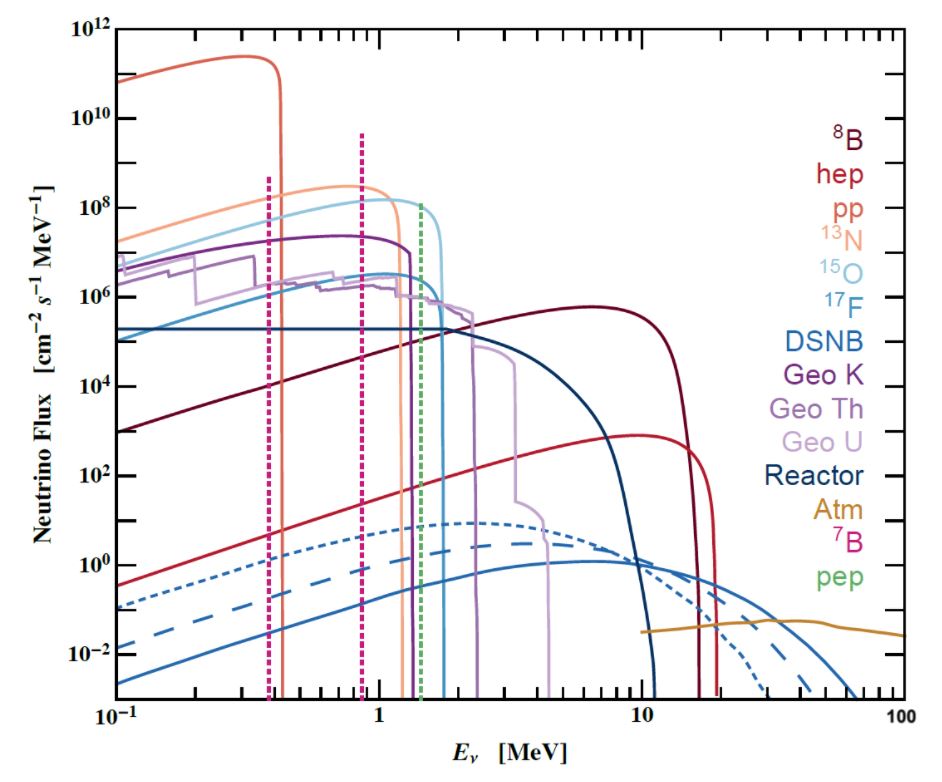} 
\end{center}
\caption{Fluxes of the dominant neutrino backgrounds for direct dark matter searches. They can be classified in broad categories as solar (pp chain and CNO cycle \cite{Robertson:2012ib}), atmospheric, diffuse supernova background, geo- and reactor neutrinos. Figure from \cite{Gelmini:2018ogy}.}
\label{fig:nu_spectra}
\end{figure*}

Figure~\ref{fig:nu_spectra} shows the expected flux of all the dominant neutrino backgrounds, assuming SNOLAB as the experiment's location. The uncertainties of these fluxes remain variable, with prospects for important improvements in the near future. Further studies are needed for addressing the impacts on DM sensitivities over a wide range of masses, especially for the backgrounds from the following astrophysical processes:
\begin{itemize}
\item \textbf{Low-energy atmospheric neutrinos}: below $\sim 100$ MeV energies, theory uncertainty is $\sim 25$\% and arise from Solar effects and geomagnetic fields \cite{Honda:1995hz,Battistoni:2005pd,Newstead:2020fie}. Extensions of atmospheric neutrino models to reach sub 100 MeV energies is ongoing \cite{Honda:2011nf}, and the Gadolinium upgrade of Super-Kamiokande \cite{Beacom:2003nk}, with first loading stage completed \cite{Super-Kamiokande:2021the}, would improve signal to noise in relevant energies. 
\item \textbf{Diffuse supernova neutrinos}: the diffuse supernova neutrino background (DSNB) uncertainty is driven by uncertainties in the cosmological rate of core collapse, the frequency of collapse to black holes, simulations of stellar core collapse, and neutrino oscillations \cite{Beacom:2010kk,Lunardini:2010ab}. Each can contribute at the tens of percent level or more. Current predictions incorporate effects such as dependence on the progenitor star and binary effects, e.g.,  \cite{Horiuchi:2017qja,Kresse:2020nto,Horiuchi:2020jnc}. The first detection of the DSNB is anticipated by gadolinium-enhanced Super-Kamiokande \cite{Super-Kamiokande:2021the}. 
\item \textbf{Solar neutrinos}: much of the solar neutrinos have been measured and the dominant $pp$ chain has percent errors while those involving heavier isotopes (7Be, 8B, and CNO) have tens of percent uncertainty. In recent years, the Standard Solar Model has not been able to explain new measurements of photospheric heavy element abundances in the Sun's atmosphere \cite{Asplund:2009fu} along with helioseismology measurements, creating a new solar ``metallicity'' problem, e.g., \cite{Villante:2013mba}.
\end{itemize}

Currently, how uncertainties should be assigned to the normalization of each of these fluxes is an open question. Ref.~\cite{Baxter:2021pqo}, taking input from all the main direct dark matter experiments, offers a possible solution and provides a set of recommended values. 

Sensitivity to neutrino backgrounds allows to exploit future DM experiments, such as argon-based ARGO~\cite{DarkSide-20k:2017zyg} or xenon-based DARWIN~\cite{DARWIN:2016hyl}, as unique instruments for exploration of neutrino astronomy and phenomenology. Taking advantage of enhanced coherent neutrino scattering, thanks to heavy targets such as argon or xenon, which is sensitive to all neutrino flavors, low keV-level thresholds, as well as good detection prospects for neutrino-electron scattering empowers large direct DM experiments to provide complementary information to conventional neutrino telescopes. 
This includes studies of solar neutrinos~\cite{Billard:2014yka,DARWIN:2020bnc}, supernova neutrinos~\cite{Lang:2016zhv}, geo-neutrinos~\cite{Gelmini:2018gqa}, pre-supernova neutrinos~\cite{Raj:2019wpy} as well as neutrinos associated with supermassive star explosions~\cite{Munoz:2021sad} that could be related to the origin of supermassive black holes. Further studies are necessary to explore the full potential as well as complementarity capabilities of upcoming direct DM detection experiments.

\subsubsection{Probes with paleo detectors}

Another handle on astrophysical backgrounds could be offered by paleo-detectors~\cite{Baum:2018tfw,Drukier:2018pdy}, a proposed alternative to conventional direct detection experiments. Paleo-detectors would use the nuclear damage tracks recorded in natural minerals over time-scales as large as a gigayear to search for sources of keV-scale nuclear recoils, promising exposures much larger than what is feasible in conventional dark matter detectors. For example, measuring the tracks in 100\,g of target material that has been recording events for 1\,Gyr corresponds to an exposure of 100\,kilotonne--years. As in conventional direct detection experiments, the signal in paleo detectors are nuclear recoils and hence, paleo-detectors could measure the same sources of astrophysical neutrino fluxes, i.e., solar neutrinos, supernova neutrinos, and atmospheric neutrinos. Unlike conventional experiments that measure nuclear recoils in real time however, paleo-detectors would measure the number of events integrated over the age of the mineral, reaching up to a billion years for minerals routinely found on Earth. This has various important implications. 

Regarding supernova neutrinos, when averaged over time-scales much longer than the few-decades average interval between core collapse supernovae in our Galaxy, the galactic supernova neutrino flux is about two orders of magnitude larger on Earth than the DSNB. Potentially, this galactic supernova neutrino flux could be measured with paleo-detectors~\cite{Baum:2019fqm}, providing a direct measurement of the galactic supernova rate. Perhaps more interestingly, using a series of paleo-detectors of different ages~\cite{Baum:2021chx}, one could obtain (coarse-grained) information about the time-dependence of various neutrino fluxes at Earth over gigayear timescales. 

An experimental program is needed to demonstrate the feasibility of paleo detectors and to better understand limitations imposed by other backgrounds, in particular natural defects in minerals that could potentially mimic nuclear recoil tracks and radiogenic backgrounds. However, if successful, paleo-detectors could allow for  measurements of the changes in the solar neutrino flux~\cite{Tapia-Arellano:2021cml}, galactic supernova rate~\cite{Baum:2019fqm}, or atmospheric neutrino flux~\cite{Jordan:2020gxx}, providing information about solar evolution, the Milky Way's star formation history, and the cosmic ray rate impinging on Earth over gigayear timescales. All of these would serve to realize new and potentially powerful methods to identify DM scattering events amongst astrophysical backgrounds~\cite{Baum:2021jak,Baum:2021chx}.

\subsection{Systematics due to dark matter distribution}
\textit{Contributors: JiJi Fan, Lina Necib, Ibles Olcina, Volodymyr Takhistov}\\

\begin{figure*}
\begin{center}
\includegraphics[width=0.8\textwidth]{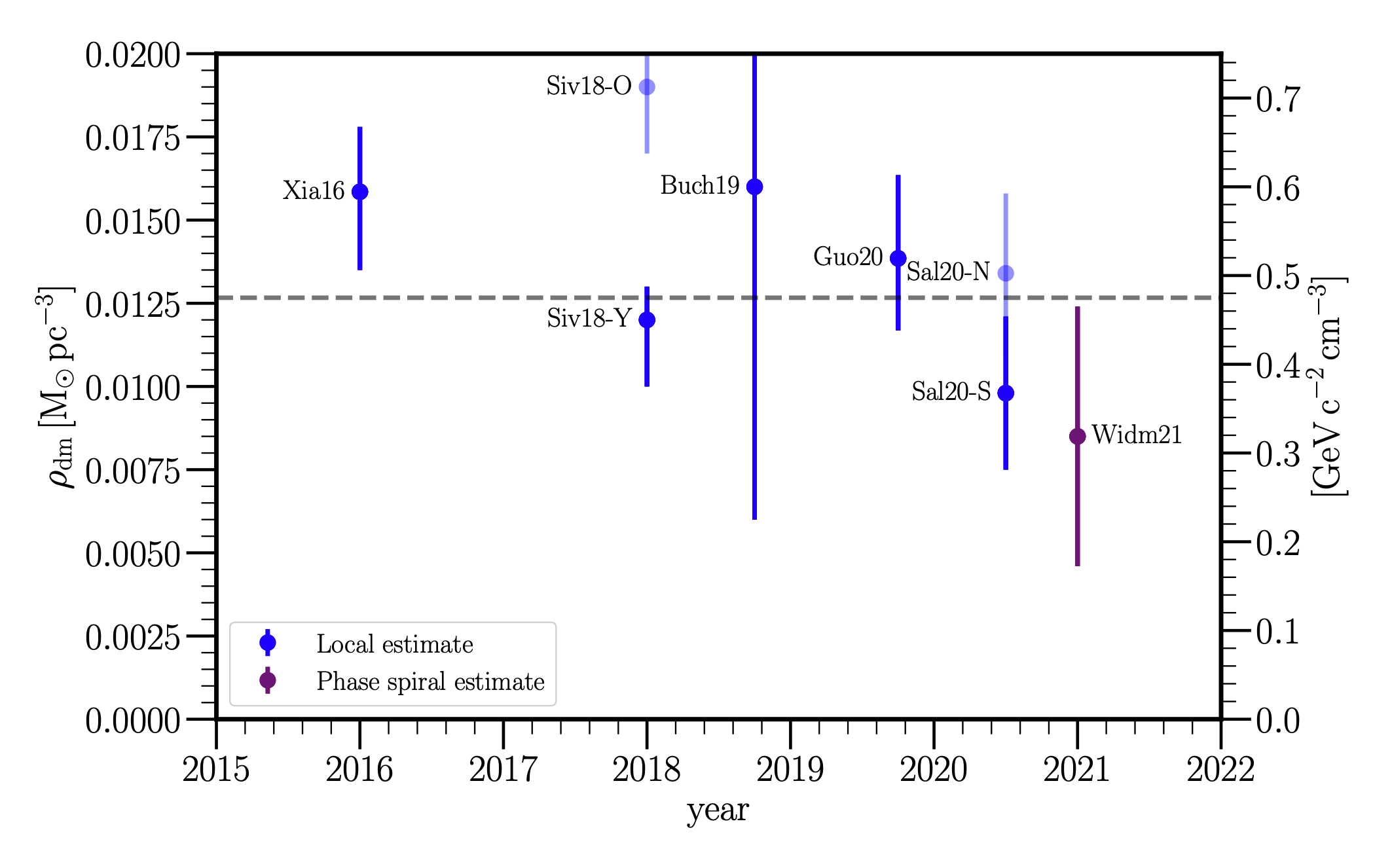} 
\end{center}
\caption{Measurements of the local density profile of DM $\rho_{\rm{DM}}$ from recent studies \cite{2016MNRAS.458.3839X,2018MNRAS.478.1677S,2019JCAP...04..026B,2020MNRAS.495.4828G,2020A&A...643A..75S,2021A&A...653A..86W}. \cite{2022arXiv220101822S} argues that these measurements, although seemingly consistent within errors, are missing effects of the tilt term in the analysis, leading to an order unity systematic error. Figure from \cite{2022arXiv220101822S}.}
\label{dm_density}
\end{figure*}

Direct detection \cite{1986PhRvD..33.3495D} of DM depends highly on the local phase space distribution of DM. More explicitly, DM direct detection rates involve an astrophysical term that contains the flux of DM through the detector, and is therefore proportional to the local DM density and velocity distribution. 

With the advances brought in by observations, particularly the \emph{Gaia} mission \cite{Gaia:2016zol,Gaia:2018ydn,2021A&A...649A...1G}, progress has been made on the DM phase space distribution. We mention some of them here:
\begin{itemize}
    \item \textbf{Local DM density distribution} Kinematic stellar catalogs have been crucial in informing the local DM density, based on the vertical motion of stars (see \cite{Read:2014qva} for a review of the methods employed to measure the local DM density). The presence of small local dark subhalos, or dark streams can affect these measurements, which introduce an error on the expected DM rate for direct detection measurements. Recent measurements have led to a value of $\rho_{\rm{DM}} \sim 0.5$ GeV/cm$^3$, as summarized in Fig.~\ref{dm_density} from Ref. \cite{2022arXiv220101822S}. However, \cite{2022arXiv220101822S} argues that the disk wobbling leads to a systematic error affecting these measurements. Future work will address these systematic errors and lead to a more accurate value of the local DM density.
    \item \textbf{Local DM velocity distribution} Direct detection experiments typically assume a Maxwell Boltzmann distribution for DM \cite{Drukier:1986tm}. However, given the large swaths of stellar structure found through \emph{Gaia} (see \cite{Helmi:2020otr} for a review), a more empirical approach is required. \cite{2019ApJ...883...27N} showed using simulations a correlation between the velocity distributions of accreted stars and accreted DM from the same mergers, building in \cite{2019ApJ...874....3N} an initial velocity distribution of the local DM from the luminous satellites. This distribution peaks at lower values than the assumed Maxwell Boltzmann distribution, leading to a suppression of the DM rate at lower masses. This is however incomplete, and further work is required to include all DM components, particularly the smaller scale local streams, and dark subhalos. Additionally, the correlation between the stellar velocities of DM and stars breaks down when non-standard interactions of the dark sector as added, and therefore need to be studied separately. Unlike the local density measurement of DM, which is a shift in the standard direct detection limit plots, differences in velocity distributions affect different experiments differently, and the most constraining experiment with a Maxwell Boltzmann velocity distribution might not actually be the one setting the strongest limits (see \cite{2019ApJ...874....3N,Buckley:2019skk,Buch:2020xyt,Buch:2019aiw}). 
    \item \textbf{DM substructure} Kinematic substructure in DM, which can be accompanied by a stellar kinematic structure like a stream, can affect direct detection rates and signal modulation \cite{Freese:1987wu}. It is therefore crucial to map out the full local kinematic history of the local solar neighborhood. This is a work in progress with the recent updates from \emph{Gaia}, (see e.g. \cite{2018MNRAS.478..611B,2018MNRAS.475.1537M,2020NatAs...4.1078N,Helmi:2020otr,2020ApJ...901...48N,Lovdal22,Ruiz22}).
\end{itemize}

Alternatively, halo-independent analysis methods for direct DM detection experiments, developed for both nucleon~(e.g.~\cite{Fox:2010bz,Gondolo:2012rs}) as well as electron~\cite{Chen:2021qao} scattering, avoid the problem of uncertain local DM distribution and instead infers its properties from data. This allows then to compare different experimental data and possible signals without making any assumptions on the uncertain local dark halo.

The impact of the DM distributions with \emph{Gaia} data on direct detection searches has just started to be explored~\cite{OHare:2018trr, OHare:2019qxc,Wu:2019nhd, Buckley:2019skk, Buch:2019aiw, Buch:2020xyt, Radick:2020qip}. Each study mainly focuses on the effects from a subset of DM distributions on either the DM-nucleon or DM-electron scattering. Further developments in understanding astrophysical DM distributions is needed to evaluate their full impacts on DM direct detection. On the other hand, future data from direct detection could potentially probe DM distributions, i.e., constrain the substructure DM fraction, even when it constitutes a sub-dominant component of the local DM density~\cite{Buch:2020xyt}.

%% file: techniques.tex
\section{Synthesis and novel techniques}

In parallel to improved understanding of astrophysical phenomena, new analysis strategies and statistical techniques also allow for powerful improvements to DM sensitivity. In this section, we cover some promising methods being explored in the literature in the context of DM search with cosmic messengers. 

\subsection{Synthesis of targets and messengers}\label{section:synthesis}
\textit{Contributors: Michael Burgess, Pat Harding, Kohta Murase, Laura Olivera-Nieto, Volodymyr Takhistov}\\

Recent coincidence observations by IceCube of neutrinos coincident with gamma-rays for blazar TXS 0506+056~\cite{IceCube:2018cha} as well as IceCube neutrinos coincident with radio signals associated with tidal disruption events ~\cite{Stein:2020xhk} have established the significance of multimessenger astronomy for uncovering the nature and mechanisms of astrophysical sources. Along with other observations, such as neutrino production from cosmic-ray line-of-sight interactions~\cite{Essey:2010er,Kochocki:2020iie}, further multimessenger studies with neutrinos will identify and assist in disentangling astrophysical background sources for indirect DM searches.

There are several open-source packages that can be used to jointly fit data from different instruments and observatories. One of them, focused on the the gamma-ray range, is Gammapy, a community developed open source Python package for gamma-ray astronomy. It relies on a common FITS-based open data format developed by an initiative called "Gamma Astro Data Formats" (GADF)~\cite{Nigro:2021xcr}. Gammapy provides methods for the analysis of high level data of many gamma-ray instruments including Imaging Atmospheric Cherenkov Telescopes (IACT), such as HESS, MAGIC and VERITAS~\cite{Nigro:2019hqf}, Water Cherenkov Observatories, such as HAWC~\cite{HAWC:2021vuj} as well as support for Fermi-LAT data. Additionally, Gammapy will be the base library for the "Science Tools" of future Cherenkov Telescope Array (CTA)\footnote{\url{https://www.cta-observatory.org/ctao-adopts-the-gammapy-software-package-for-science-analysis/}}. Data can be combined at different stages, ranging between relatively low-level data products like event lists and instrument response functions, but also higher level data products like flux points, if the corresponding likelihood profiles are provided as well. This is relevant for stacked analysis, for example. 

Another high-level open code that can be used to combined experiments in multi-messenger or multi-wavelength analysis is the Multi-Mission Maximum Likelihood framework (3ML)~\cite{Vianello:2015wwa}. Rather than unifying the underlying base analysis of each experiment, each experiment has a written plugin which takes model input and gives output of likelihood to the main 3ML framework. In this way, the under-the-hood behavior of each experiment is left to that experiment itself, without any enforced formats, modules, or capabilities - as long as the experimental code takes in a model and outputs a likelihood, it can be added into a joint analysis. In this way, 3ML is always able to use the best possible up-to-date algorithms and data from each experiment and maintained by that experiment as it best sees fit. The nature of the 3ML code is such that it is outside the space of individual experiments, although it is regularly used for analysis with the HAWC observatory\footnote{\url{https://threeml.readthedocs.io/en/stable/notebooks/hal_example.html}}. The 3ML framework itself is responsible for the combination of likelihoods from the input experiments as well as the minimization, Markov Chain Monte Carlo, or nested sampling used to derive best-fit model parameters and confidence intervals. To be as general as possible, 3ML has a wide code base of model examples and templates to consider for any analysis.

Independent on the specific package, the ongoing development of open-source tools and the definition of standards that facilitate the sharing of data will be crucial to fully exploit the synergies between different wavelengths and astrophysical messengers.

\subsection{Novel techniques}
\label{sec:novel_techniques}
\textit{Contributors: Shin'ichiro Ando, Ilias Cholis, Shunsaku Horiuchi, Nick Rodd}\\

A fundamental problem in astronomy affecting indirect DM search is the detection of point sources. When studying a population of sources, as the individual sources dim, there is a transition to a regime where characterizing the properties of any one object becomes unreliable, and instead one must study the features of the population as a whole. These novel statistical techniques take us beyond the energy spectrum of the diffuse emissions, including spatial distributions and cross correlations with source catalogs.

\subsubsection{Angular power spectrum}

If the photon data are dominated by a handful of bright point sources rather than by infinite number of dim sources, statistics -- especially the flux distribution -- of the photons will be very characteristic.
One of the simplest statistics is the angular power spectrum that incorporates the variance of the flux distribution.
Analyzing the modern gamma-ray data sets using the angular power spectrum was proposed in order to distinguish various astrophysical point sources and dark matter annihilation or decay~\cite{Ando:2005xg, Ando:2006cr}.
The angular power spectrum of the all-sky gamma-ray background data obtained with the Fermi-LAT was detected in 2012~\cite{Fermi-LAT:2012pez}, which revealed dominant contribution from the gamma-ray blazars~\cite{Fornasa:2016ohl, Fermi-LAT:2018udj}.
Neither intrinsic source clustering nor signatures of particle dark matter that might be encripted in the angular power spectrum, however, has not been detected yet.

\subsubsection{Non-Poissonian template fit}

One of the most widely used methods for studying point-source populations is the statistical framework of the non-Poissonian template fit (NPTF)~\cite{Malyshev:2011zi,Lee:2014mza,Lee:2015fea,Mishra-Sharma:2016gis}.
The NPTF rose to prominence given its application to the question of whether the anomalous Galactic Center $\gamma$-ray emission was associated with a population of point-sources, or instead was a signal of dark-matter annihilation, although it has also been applied to neutrino datasets~\cite{IceCube:2019xiu}.
The method exploits the fact that a population of sources can predictably modify the observed distribution of photons away from the Poisson distribution expected of smoother emission -- as encapsulated in an analytic likelihood.

In order to be computationally tractable, the NPTF makes approximations, such as assuming each pixel in a binned sky-map is statistically independent, even though they are demonstrably not (the finite point spread function of any instrument can correlate the pixels around a point source).
More recently, it has been shown that NPTF suffers from biases that lead it to generate spurious evidence for point sources~\cite{Leane:2019xiy,Leane:2020nmi,Leane:2020pfc}.
A subset of the issues identified can be resolved~\cite{Buschmann:2020adf,Chang:2019ars}, however, others appear more fundamental.

The identification of shortcomings in the NPTF has spurred recent efforts to introduce more reliable methods.
The Compound Poisson Generator (CPG) framework introduces a novel likelihood that removes a number of biases fundamental in the NPTF framework, leading to improved performance particularly for X-ray point-source searches~\cite{Collin:2021ufc}.
However, even this method assumes nearby pixels are independent.
With a view to accounting for these missed correlations, several groups have pursued an approach based on convolutional neural networks~\cite{Caron:2017udl,List:2020mzd,List:2021aer,Mishra-Sharma:2021oxe}.
These methods appear to be able to avoid the pitfalls of the NPTF, and further exhibit considerable sensitivity to dim sources.
However, going forward, these machine learning based approaches will need to be tested considerably to determine whether their results are robust, or whether they have their own set of challenging systematics.

Refined versions of these methods will have an important role to play in characterizing any emerging indirect detection dark-matter signal.
Point sources are ubiquitous, and validating that any putative signal of new physics is not a dim population of sources will be an important step in building confidence on the path to discovery.
A clear example of this is provided by the Galactic Center excess: the possibility that this is in fact a signal of annihilating dark-matter has yet to be robustly excluded, although there are by now a several of strains of evidence disfavoring the possibility.
Yet the leading alternative hypothesis to dark matter remains point sources; the debate around this excess -- and indeed future anomalies -- may eventually be resolved by the successors to the NPTF. More details can be found in an accompanying whitepaper \cite{SnowmassCF1WP6}. 

\subsubsection{Wavelets}

Wavelets are mathematical objects that have been used widely in image analysis at various wavelengths to decompose images in structures of different scales \cite{1992tlw..conf.....D, 2006Stack}.
Thus, wavelets rely less on the presumptive distinction of backgrounds vs signals and may provide an alternative way forward. It has been shown that wavelets can detect pixels where point sources bellow Fermi detection threshold  may lie \cite{Bartels:2015aea, Zhong:2019ycb}. Such techniques work best on high-resolution maps that impose a small angular scale cut-off only at angles smaller than the point spread function. Moreover, they require high statistics and an extensive search over the best type of wavelet family of functions used. Future analyses will greatly benefit from the continuous \textit{Fermi}-LAT observations.

\subsubsection{Cross correlations}

Cross correlating different datasets allows correlated information to be extracted while removing uncorrelated noise. For example, if gamma-ray emission from astrophysical sources are correlated with emission in other low-frequency wavebands (e.g., optical, radio, etc.), one can fully utilize the correlated information in order to study the gamma-ray source properties or extract information on particle DM. 
Similarly, since most, if not all, astrophysical sources trace the large-scale structure of DM, one can consider correlating with datasets with tracers of large-scale structure such as catalogs of galaxies or gravitational lensing maps. 

If a catalog of galaxies focuses on a particular redshift range, taking the cross correlation with that catalog will essentially single out source contributions from the same redshift range, and at the same time remove source contributions from other redshift ranges~\cite{Ando:2014aoa}. This \textit{tomography} provides powerful diagnostics for searchiing for DM and astrophysics alike. 
Since the contributions of DM annihilation or decay are predicted to be biased towards lower redshift ranges when compared with ordinary astrophysical sources such as such as blazars and starburst galaxies, taking the cross correlations with nearby galaxy catalogs such as 2MASS is optimal for the search of particle DM~\cite{Ando:2013xwa}.
On the other hand, one must be mindful that galaxies are a biased tracer of the DM distribution; cross correlating with weak gravitational lensing catalogs has the advantage that they are an unbiased tracer of the DM distribution, which makes the technique complementary to the one with galaxy catalogs~\cite{Camera:2012cj,Camera:2014rja}.

Both of these strategies have been performed with the Fermi-LAT data. Cross correlation with various galaxy catalogs have showed positive correlations in many cases~\cite{Cuoco:2015rfa, Cuoco:2017bpv}. Cross correlation with lensing maps have been limited by lensing coverage \cite{Shirasaki:2014noa,Troster:2016sgf,Shirasaki:2018dkz}, but recently, the Dark Energy Survey (DES) collaboration found possible cross correlation between the Fermi-LAT gamma-ray data and DES weak lensing data, which is found consistent with theoretical expectation with gamma-ray blazar models~\cite{DES:2019ucp}. 

In the future, cross correlation is a technique which is anticipated to become even more powerful as gravitational tracers become more complete. For example, the Vera C. Rubin Observatory will increase sky coverage of existing lensing maps by some factor $\sim10$ or more, contributing to an significant improvement in the amount of data that can be correlated. In parallel, as the properties of astrophysical sources are better understood---through multi-messenger astrophysics including the cross correlation technique---the information will further feedback into future searches for DM.

%% file: conclusion.tex
\section{Concluding remarks}
The coming decade of DM searches will be driven by both strong sensitivity improvements with new instruments, synergies between experiments, as well as the rise in multi-messenger astrophysics. By coming together and sharing data, DM studies in direct detection, indirect detection, and natural laboratories can all benefit from each other, and furthermore all can benefit from the progress in non-DM-driven studies from multi-messenger astrophysics. In this whitepaper, we have covered major backgrounds to DM searches with photons, cosmic rays, neutrinos, as well as gravitational waves; and highlighted important future developments of astrophysical sources and processes, as well as new numerical techniques to merge datasets. With synergistic studies, the community can get the most science out of its data and can delve even deeper into the DM parameter space. The future is promising for the search for DM and by continuing to work together as a community we can strengthen it further. 